\documentclass{appolb}
\usepackage{epsfig}
\usepackage{cite}
\usepackage{tcolorbox}
\tcbuselibrary{theorems}
\def\sr17{$\sqrt{s}$~=~17~GeV~}

\newcommand{\be}{\begin{equation}}
\newcommand{\ee}{\end{equation}}

\begin{document}
%\date{\today}
%\pagestyle{plain}
%% uncomment the following line to get equations numbered by (sec.num)
%\eqsec
\newcount\eLiNe\eLiNe=\inputlineno\advance\eLiNe by -1
\title{Hadronic Resonance Gas Model and  Multiplicity Dependence in 
 p-p, p-Pb, Pb-Pb collisions:\\
Strangeness Enhancement
\thanks{Presented at {\it Criticality in QCD and the Hadron Resonance Gas}, University of Wroc\l aw,  Poland , July 2020}%
}
	\author{J.~Cleymans%
	\address{UCT-CERN Research Centre and Department  of  Physics,\\ University of Cape Town, Rondebosch 7701, South Africa}
\and
	Pok Man Lo
	\address{Institute of Theoretical Physics, University of Wroc\l aw, PL-45204, Poland}
\and	
K.~Redlich
	\address{Institute of Theoretical Physics, University of Wroc\l aw, PL-45204, Poland}
	\and
	N.~Sharma
	\address{Department of Physics, Panjab University, Chandigarh 160014, India}
}
\maketitle
\begin{abstract}
%Recent theoretical work on the multiplicity dependence of strange hadrons is presented. 
Recently,  the ALICE collaboration has observed an interesting systematic behavior of  ratios of 
identified particles to pions yields at the LHC,  showing that they   depend solely on the  charged-particle multiplicity $dN_{ch}/d\eta$, in p-p,   p-Pb and Pb–Pb collisions. 
%In particular, the evolution of  strange  particle yields  with $dN_{ch}/d\eta$ shows  patterns that are common  in p-p,   p-Pb and Pb–Pb collisions. 
In particular, the  yields of (multi)strange  particles, relative to pions increases  with $dN_{ch}/d\eta$ and the enhancement  becomes more pronounced with increasing strangeness content. We will argue, that such a pattern of strangeness enhancement is arising naturally in the thermal model accounting for  exact strangeness conservation. Furthermore,  extending the thermal model by including hadron interactions within the S-matrix approach, the ALICE data can be well  quantified by the thermal particle yields at  the chiral-crossover temperature,  as previously found in central Pb-Pb collisions.  
\end{abstract}
%%%%%%%%%%%%%%%%%%%%%%%%%%%%%%%%%%%%%%%%%%%%%%%%%%%%%%%%%
\section{Introduction}
%%%%%%%%%%%%%%%%%%%%%%%%%%%%%%%%%%%%%%%%%%%%%%%%%%%%%%%%%
The thermal description of hadronic yields measured in nucleus-nucleus collisions by  using the hadron resonance gas model (HRG)  has been already very successful, see e.g.~\cite{Andronic:2017pug,Andronic:2018qqt}.
The identifying feature of the HRG  is that all stable hadrons and known hadronic resonances  listed
in the review of particle physics~\cite{Tanabashi:2018oca} are
assumed to be in thermal and chemical equilibrium.
This  assumption drastically reduces the number of free parameters,  as this stage is determined by just  %a  few
%thermodynamic variables namely,
the chemical freeze-out temperature $T$, the various chemical potentials $\vec{\mu}$ determined by
the conserved quantum numbers,  and by the volume $V$ of the system.
It has been shown that this description is  valid  not only for a static, but also for  an expanding fireball  
~\cite{Cleymans:1999st,Broniowski:2001we,Akkelin:2001wv} that follows  Bjorken's  scaling expansion~\cite{Bjorken:1982qr}.
For an expanding system, and  after integration over transverse momenta of produced hadrons, % $p_T$ these authors have shown that:
%
%
%\begin{equation}
%\frac{dN_i/dy}{dN_j/dy} = \frac{N_i^0}{N_j^0}
%\label{eq:bjorken}
%\end{equation}
%where $N^0_i$ ($N_j^0$)is the particle yield of hadron $i$ ($j$)
% as calculated in a fireball at rest, while $dN_i/dy$ is the yield of  hadron $i$  on the rapidity plateau.
%Hence, in the Bjorken model with longitudinal scaling and radial expansion 
the effects of hydrodynamic flow cancel out in yields ratios, resulting in the same values as calculated in a fireball at rest.
Furthermore, the HRG was also shown to provide a parameter free description of the equation  of state and some fluctuation observable calculated in lattice QCD (LQCD) in the hadronic phase\cite{karsch,Goswami:2020yez}.  The above provides strong evidence  that HRG statistical operator is a very good  approximation of the QCD partition function, and that particle  yields produced in heavy-ion collisions are consistent with the expectation of first-principles based  LQCD calculations.  

The yields produced in heavy-ion collisions have been the subject of intense discussions over the past few years
and several proposals have been made  to further improve the  HRG model. This was particularly  in view of the fact,  that in the   HRG the number of pions was  underestimated while the number of protons was  overestimated. Several proposals to further improve or extend the HRG have been made recently:
\begin{itemize}
\item incomplete hadron mass-spectrum~\cite{Goswami:2020yez,Noronha-Hostler:2014aia,marczenko,new},
\item chemical non-equilibrium at freeze-out~\cite{Petran:2013lja,Begun:2013nga,Begun:2014rsa},
\item modification of hadron abundances in the hadronic phase~\cite{Steinheimer:2012rd,Becattini:2012xb,Becattini:2016xct},
\item separate freeze-out for strange and non-strange hadrons~\cite{Chatterjee:2013yga,Bellwied:2013cta,Chatterjee:2016cog},
\item excluded volume interactions~\cite{Andronic:2017pug,Andronic:2018qqt,Alba:2016hwx},
\item  energy dependent Breit-Wigner widths~\cite{Vovchenko:2018fmh},
\item use the phase shift analysis to take into account repulsive and attractive interactions~\cite{Lo:2015cca,Dash:2018mep,Andronic:2018qqt},
\item use the K-matrix formalism to take interactions into account~\cite{Dash:2018can}.
\end{itemize}
%These proposals improve the agreement with the observed yields and furthermore, some  of them  change the chemical freeze-out
%temperature, $T_{ch}$ in only a minimal way like those presented recently in~\cite{Vovchenko:2018fmh,Andronic:2018qqt}.
%%%%%%%%%%%%%%%%%%%%%%%%%%%%%%%%%%%%%%%%%%%%%%%%%%%%%%%%%%
\begin{figure*}[ht]
\centering
\includegraphics[width=0.9\linewidth,height=10cm]{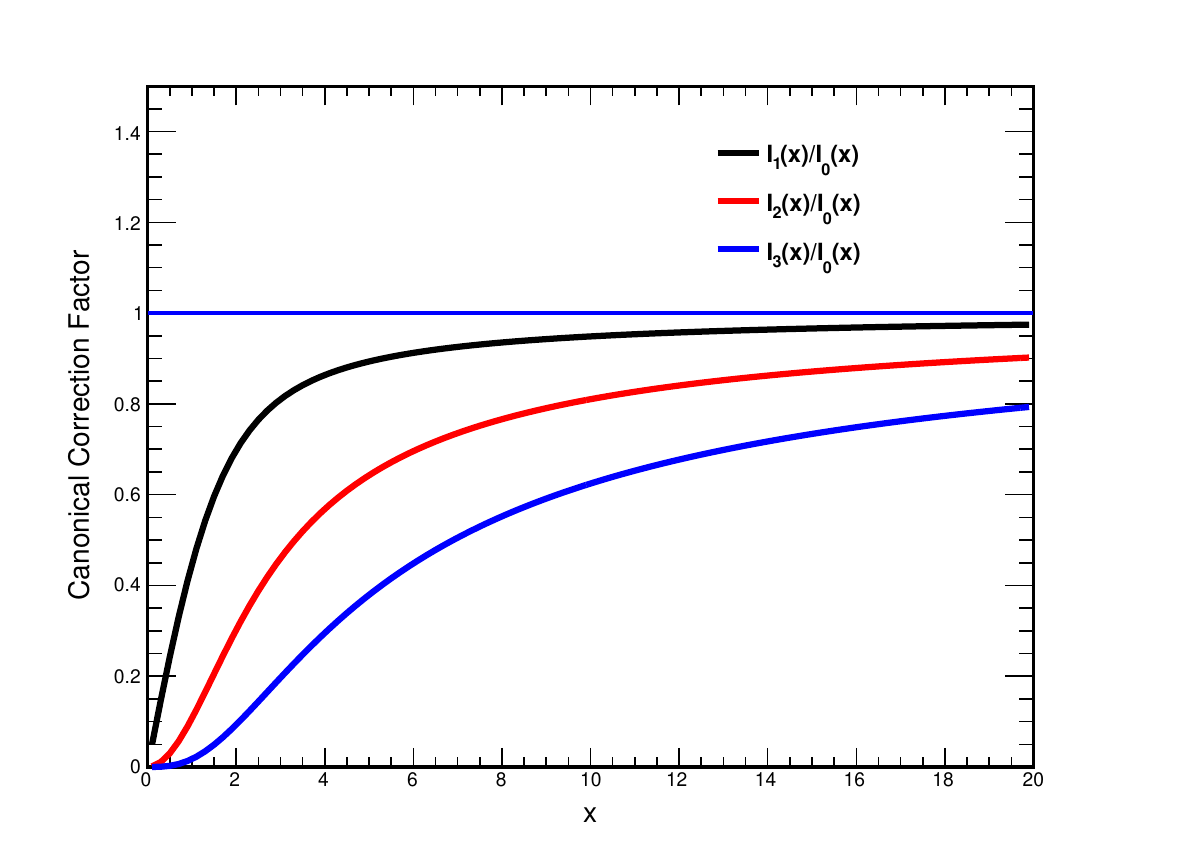}
\label{bessel}
\vskip -0.5cm
\caption{Ratio of Bessel functions relevant for determining canonical corrections to particle multiplicities.}
\end{figure*}
%%%%%%%%%%%%%%%%%%%%%%%%%%%%%%%%%%%%%%%%%%%%%%%%%%%%%%%%%%
We  argue that  in view of the recent ALICE collaboration data, in particular, the observed $dN_{ch}/dy$ dependence of particle yields 
 of the (multi)strange hadrons in pp, pA and AA collisions at the LHC \cite{ALICE:2017jyt}, essential improvements  of the
 HRG model taking into account exact strangeness conservation and hadron interactions ~\cite{Cleymans:2020fsc} can be made.  

 In the
present analysis
we  keep the basic structure of the HRG as determined in central Pb-Pb collisions
with a single chemical freeze-out temperature, $T \approx 156.5 $ MeV  and with the  off-equilibrium  suppression factor  $\gamma_s =1$, as well as  with all chemical potentials set to zero \cite{Andronic:2017pug}.
The observed  deviations  at low multiplicities will be ascribed to imposing strangeness conservation via the canonical
ensemble as explained in detail below.  The deviations seen in the yields of protons and $\Lambda$'s is   corrected by
including interactions via the S-matrix formalism~\cite{Dashen:1969ep}. These lead to a 25\% reduction in the proton yield
and a 23\% enhancement of the $\Lambda$ yield~\cite{Cleymans:2020fsc}.

We will show, that a very good description is obtained for the variation of the strangeness content in
the final state as a function of the
number of charged hadrons at  mid-rapidity  by   using the same 
fixed temperature value as in central Pb-Pb collisions, whereas  
 the volume of the fireball at mid-rapidity is found to be linearly dependent on the number of charged hadrons.  
 For small multiplicities,   the canonical ensemble  with global  strangeness conservation is needed to quantify 
the observed suppression of 
 (multi-)strange baryons. 
This is obtained  by introducing a  conservation of strangeness in the whole phase-space which is parameterized  by the  canonical correlation volume that is  larger than the  fireball volume at  mid-rapidity. 
The interactions introduced by the phase shift
analysis via the S-matrix formalism are essential for a quantitative  
description of the yields data. A more detailed description of the model and its 
comparison    with ALICE data can be found in  \cite{Cleymans:2020fsc}.

\section{Including hadron  interactions by using empirical  phase shifts}
The change in the particle yield due to interactions is taken into account by introducing the kernel function $B(M)$, which is linked to scattering   phase shifts as follows~\cite{Friman:2015zua}
\begin{equation}
B(M) = 2\frac{d}{dM}\delta(M)  ,
\end{equation}
where $\delta(M)$ is given by the empirical phase shift analysis.
For a well-defined resonance this can be replaced by~\cite{Friman:2015zua}
\begin{equation}\label{width}
B(M)\approx 4\frac{M^2\Gamma_R}{(M^2 - M_R^2)^2 + M^2\Gamma_R^2},
\end{equation}
where $\Gamma_R$ is the width of the resonance and $M_R$ is its mass.
Finally, for a narrow resonance this can be approximated by:
\begin{equation}\label{delta}
B(M)\approx    4\pi M \delta(M^2-M_R^2).
\end{equation}
The particle yields are modified by an integral over the kernel  function $B(M)$ as follows \cite{Dash:2018mep,Lo:2017ldt,Friman:2015zua}:
\begin{equation}\label{smatrix}
N(T,M) =  \int_{m_{th}}^\infty \frac{dM}{2\pi} B(M) N^{id}(T,M)
\end{equation}
where $N^{id}$ is the particle density given by the ideal gas formula.
From Eqs. \ref{width} and \ref{smatrix} it is clear, that for a system with purely attractive  interactions which are 
dominated by the formation of  resonances, the yields of particles are  as in the mixture of ideal gases of stable hadrons and 
resonances,  what  constitutes the frequently used HRG model. 
%%%%%%%%%%%%%%%%%%%%%%%%%%%%%%%%%%%%%%%%%%%%%%%%%%%%%%%%%

The extension of the HRG model as in Eq. \ref{smatrix},   and an analysis using the available phase shift data has 
been performed in~\cite{Cleymans:2020fsc,Lo:2017ldt,Friman:2015zua}. In the following we put particular attention to such 
a model  extension that accounts for interactions of nucleons and hyperons to discuss the observed production 
of (multi)strange baryons  in different colliding systems at the LHC \cite{Cleymans:2020fsc}. 

%%%%%%%%%%%%%%%%%%%%%%%%%%%%%%%%%%%%%%%%%%%%%%%%%%%%%%%%%

%{Including interactions using  phase shifts}~\cite{Lo:2017ldt}
%{Phase Shifts: correction factor of about 25\% for protons}
%\begin{center}
%\includegraphics[width=\linewidth,height=7.0cm]{fig_1a.pdf}
%\end{center}
%
%%%%%%%%%%%%%%%%%%%%%%%%%%%%%%%%%%%%%%%%%%%%%%%%%%%%%%%%%
%
%{Phase Shifts}~\cite{Lo:2017ldt}
%\begin{center}
%\includegraphics[width=\linewidth,height=7.0cm]{fig_1b.pdf}
%\end{center}
%
\begin{figure*}[ht]
\centering
\includegraphics[width=0.9\linewidth,height=10cm]{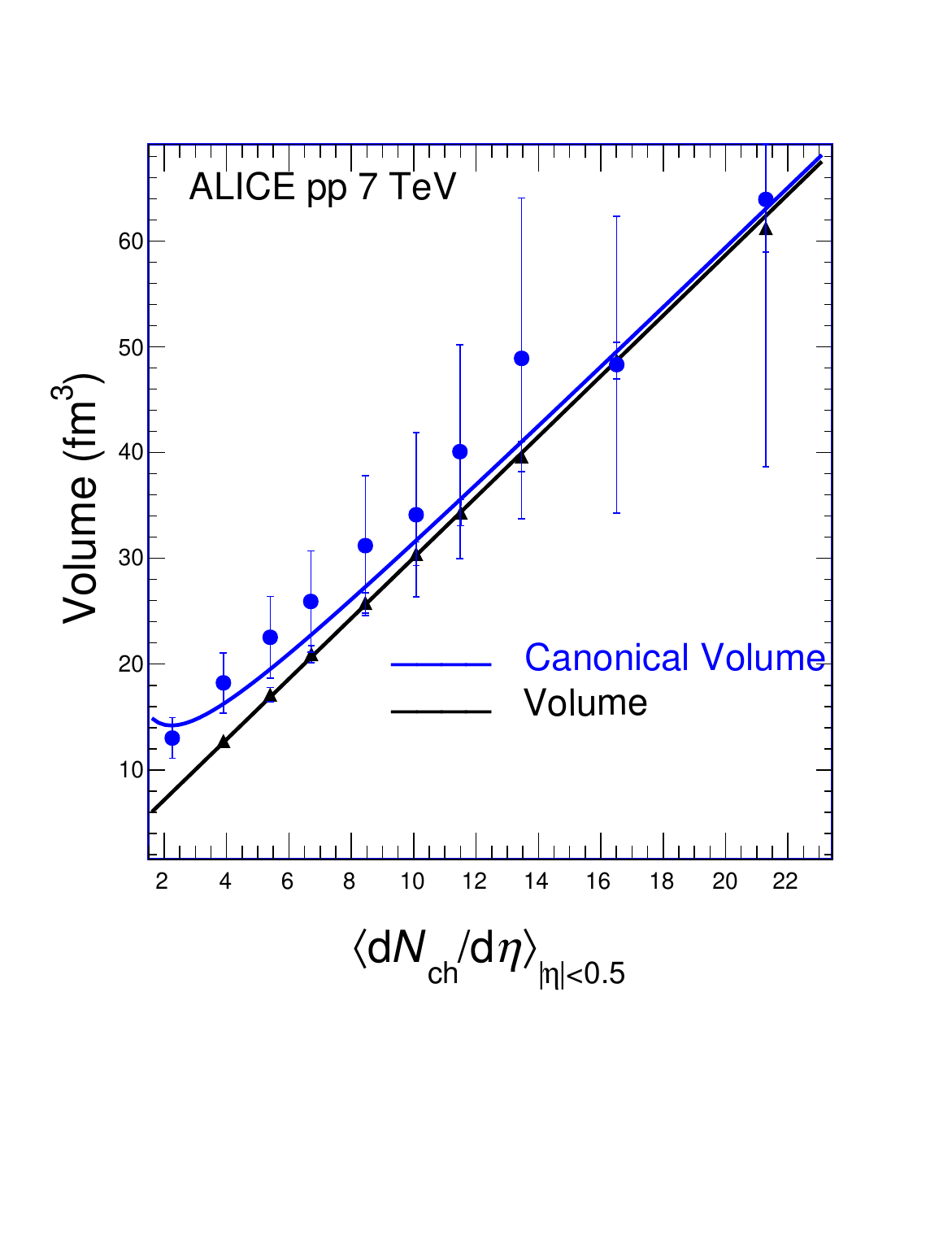}
\label{volume}
\vskip -2.0cm
\caption{Volume determined, using the HRG, from the ALICE collaboration~\cite{ALICE:2017jyt} for pp collisions at 7 TeV.}
\end{figure*}

\begin{figure*}[ht]
\centering
\includegraphics[width=\linewidth,height=9cm]{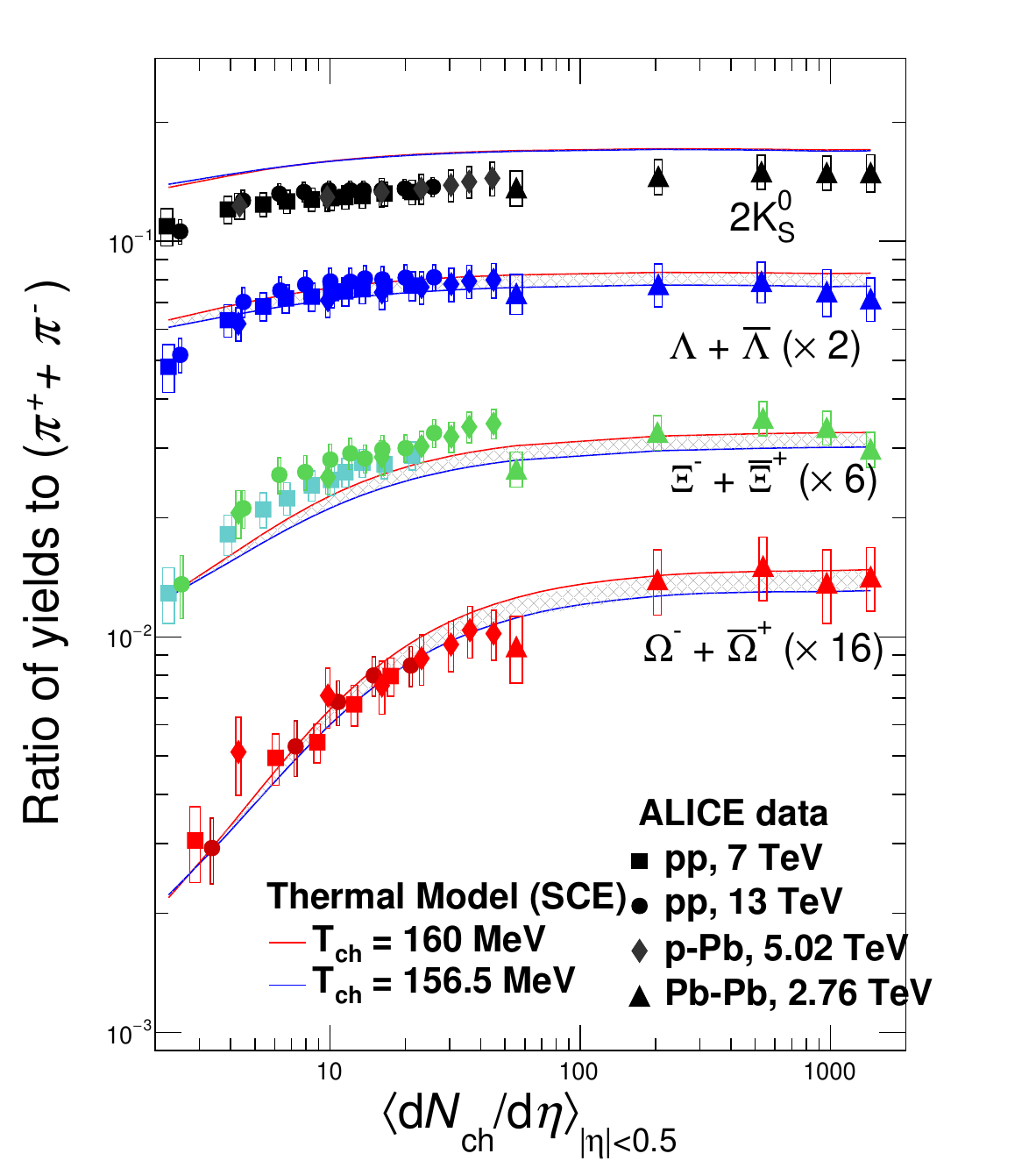}
\label{RatiosOfYields}
\caption{Ratios of yields of strange particles to pions versus charged particle multiplicity.}
\end{figure*}
%%%%%%%%%%%%%%%%%%%%%%%%%%%%%%%%%%%%%%%%%%%%%%%%%%%%%%%%%%
\section{Strangeness Canonical Ensemble}
%%%%%%%%%%%%%%%%%%%%%%%%%%%%%%%%%%%%%%%%%%%%%%%%%%%%%%%%%%
It is by now a  well established fact that the multiplicity of charged particles in the grand canonical and canonical ensembles, formulated with respect to conservation laws, can differ substantially. Indeed, 
it was pointed out by Hagedorn~\cite{hagedorn} %some fifty years ago 
that thermal models can  overestimate
the yield of charged particles %strange particles 
when the grand
canonical ensemble is used. Applying  Hagedorn's  argument to strangeness conservation, the reason for this is that when the number of
particles as well as the interaction volume are small one has to take into account
that strange particles  must be balanced by  anti-strange particles to exactly conserve strangeness. 
Thus, the abundance of e.g. $K^+$ in the fireball of volume $V_A$  at temperature $T$, will not be proportional to the
standard Boltzmann factor given by:

\begin{equation}
N_{K^+} \approx V_A\exp\left(-\frac{m_{K^+}}{T}\right)
\label{eq:boltzmann}
\end{equation}
but instead by:
\begin{eqnarray}
N_{K^+} &\approx& V_Ae^{-\frac{m_{K^+}}{T}} V_C
\left[g_{\bar{K}}\int \frac{d^3p}{(2\pi)^3}e^{-\frac{E_{\bar{K}}}{T}} \right.
 + \left. g_\Lambda \int \frac{d^3p}{(2\pi)^3}e^{ -\frac{E_\Lambda}{T}+\frac{\mu_B}{T}}+..\right]
\label{eq:boltzmann_hagedorn}
\end{eqnarray}
This introduces one more  exponential factor and implies a strong suppression of  $K^+$ yields  and leads to a quadratic   volume 
dependence, where    $V_C$ is introduced as the canonical   volume   where  exact strangeness conservation  is fulfilled. 
In  general,  $V_A\neq V_C$. 

% new  volume  which we will refer to as canonical volume dependence.

The inclusion of constraints of exact strangeness conservation in this
framework has been considered at a very early
stage, see e.g.~\cite{Redlich:1979bf}. The partition function is modified by including a $\delta$-function to enforce
strangeness to be exactly zero:
\begin{align}
 \mathrm{ Z_{S=0} = Tr \left( e^{-(E-\mu)/T} \delta_{S,0}\right)} .
\end{align}
This leads to  replacing the standard grand canonical expression, e.g. for kaons
\begin{align}
N_K = V_A e^{\frac{\mu}{T}} \int \frac{d^3p}{(2\pi)^3} e^{-\frac{E}{T}}
\end{align}
by the following
\begin{align}\label{ce}
N_K = V_AF_S \int \frac{d^3p}{(2\pi)^3} e^{-\frac{E}{T}}
\end{align}
where
\begin{align}
F_S = \frac{I_1(x)}{I_0(x)}\frac{S_1}{\sqrt{S_1S_{-1}}},
\end{align}
and $x \equiv 2\sqrt{S_1S_{-1}}$ with  $S_1 = Z_{\bar{K}} + Z_\Lambda + ...$ i.e., 
 $S_{s=\pm 1} =\sum_k Z_{k,s}$,  where the sum is  taken  over all particles and
resonances that carry   strangeness $s$.   The one-particle  partition function,   
$Z_{k,s}=V_C n_{k,s}(T)$,   with the particle  density 
\begin{align}
n_{k,s}(T)  = g_{k,s}\int \frac{d^3p}{(2\pi)^3} e^{-\frac{E_{k,s}}{T}}.
\end{align}
In view of $\vec \mu=\vec{0}$ the thermal phase space of particles is the same as for antiparticles, i.e. $S_s=S_{-s}$. Consequently, from Eq. \ref{ce}  the yields of kaons is obtained, as 
%is the sum of one-particle partition functions.
%The correction factor $S$  depends on the volume $V$ which is not necessarily the same, it will be
%referred to as the canonical volume $V_C$ and correspondingly a canonical radius $R_C$.
%For $\mu_B = 0$, as is relevant at LHC energies, one replaces the standard expression
%\begin{align}
%N_K = V \int \frac{d^3p}{(2\pi)^3} e^{-\frac{E}{T}},
%\end{align}
%with
\begin{align}
N_K = V_A\frac{I_1(x)}{I_0(x)} \int \frac{d^3p}{(2\pi)^3} e^{-\frac{E}{T}}
\end{align}
%and $x \equiv Z_{\bar{K}} + Z_\Lambda + ...$. 
For small values of $x$ this leads to the correction factors
given in Eq.~(\ref{eq:boltzmann_hagedorn}), whereas for large $x>>1$ it reproduces the grand canonical result. 

The ratio of Bessel functions, $F_S=I_1/I_0$  is  shown in Fig.~\ref{bessel}.  It is clear, that  $F_S$  starts from zero and 
 approaches one for large
values of the argument. Thus,  corrections due to exact strangeness conservation reduce particle yields,  and 
 disappear for large values of $x$, i.e. for large $T$ and or $V_C$ where the yields of particles  reach their grand canonical ensemble  
values.  % or, for large values of $S_1$ defined above.

The canonical formalism summarised above can be also generalized to thermal systems which contain contributions from multistrange particles. 
In this case the yield of particle carrying strange quantum number s, can be approximated by: 
\begin{align}
\label{equ10}
\langle N_s\rangle_A \simeq V_A \, n_s \, \frac{I_{s}(S_1)}{I_{0}(S_1)}.
\end{align}
Thus, the ratio  ${I_{s}(S_1)/ I_{0}(S_1)}$, similarly to $F_S$ in Eq. \ref{ce},  is just a suppression factor which  decreases in magnitude   with increasing  $s$ of  hadrons  and with decreasing thermal phase-space occupied by strange particles as described by the argument  $S_1$  of the Bessel functions.
Relevant ratios are shown in Fig. 1.  
A decrease of $S_1=V_C\sum_k n_{k,{s=1}}$ is  due to decreasing $T$,   or decreasing $V_C$.  These are the main properties of strangeness canonical suppression that have been introduced  \cite{Hamieh:2000tk} to describe  thermal production of multi(strange) hadrons   in heavy-ion collisions.

\section{Strangeness production at the LHC}
In the following we will show that the thermal HRG model, that 
includes S-matrix corrections for baryon interactions and accounts for exact strangeness conservation,
 provides a very good description and understanding of 
multistrange particle production systematic as observed by the ALICE collaboration at the LHC in pp, pA and AA collisions.    

In the application of the HRG model to particle production in heavy ion collisions the temperature and 
baryon-chemical  potential at  freezeout are linked to the collisions energy whereas the volume parameters to charged particle multiplicities. At the LHC, the  particles and antiparticles are produced with the same abundance, thus the thermal parameters characterising yields of particles are only the temperature and the volume. Furthermore, due to the observed coincidence of chemical freezeout temperature in the most central AA collisions and  the chiral-crossover  temperature obtained in lattice QCD  we assume the same $T=156.5$ MeV for all colliding systems   at fixed $\sqrt s$.  Consequently, to quantify strange particle yields and their dependence on $dN_{ch}/dy$  as observed by ALICE collaboration at the LHC,  within  the above thermal model there are only two volume parameters to be extracted from the data. 
The fireball volume $V_A$ which is  obtained from  fits to measured yields of pions and protons at mid-rapidity, and the canonical volume $V_C$ that is fitted to reproduce yields of strange and multistage hadrons.

The above thermal  analysis for each multiplicity bin, for p-p, p-Pb and Pb-Pb has been done
 using the latest version of the THERMUS code~\cite{Wheaton:2004qb}~\footnote{B. Hippolyte and Y. Schutz,
https://github.com/thermus-project/THERMUS}, that was extended to include S-matrix corrections for baryon interactions.
The extracted volume parameters are shown in Fig.~\ref{volume}. 
As can be seen, the fireball volume can be determined very accurately with linear dependence on $dN_{ch}/dy$, whereas 
  the canonical volume cannot, especially  for
large multiplicities where within errors  it is consistent
with being equal to the fireball volume.
For low multiplicities, however,  there is  a clear difference, indicating that $V_C>V_A.$ The dependence of  strange  particle ratios as a function of multiplicity
 seen in Fig.~\ref{RatiosOfYields} follows the  behavior
predicted in~\cite{Hamieh:2000tk,Redlich:2001kb}. Furthermore data  can be quantified, with 
the above formulation of the  HRG model, quite satisfactorily.  
The model  prediction on different particle yields  agrees with  the data up to  two standard deviations  for  all $dN_{ch}/d\eta$. 
The data on pion yields  are always slightly above the calculated points while the kaons are always below. This has implications for the kaon to pion
ratio which is seen in Fig.~\ref{RatiosOfYields} to exhibit  the largest deviations from the data, illustrating   the pitfalls of comparing  ratios in the thermal model. It is better to compare directly yields.
A more  complete  description of the model setup and its comparison with  ALICE data can be found in ~\cite{Cleymans:2020fsc}.

%%%%%%%%%%%%%%%%%%%%%%%%%%%%%%%%%%%%%%%%%%%%%%%%%%%%%%%%%%
\section{Conclusions}
%%%%%%%%%%%%%%%%%%%%%%%%%%%%%%%%%%%%%%%%%%%%%%%%%%%%%%%%%%%
We have shown that the observed behavior of (multi)strange hadron yields in pp, pA and AA collisions at the  LHC  with  charged particle multiplicity can be explained  naturally in the thermal model which accounts for exact strangeness conservation and S-matrix corrections to baryon interactions. The freezeout temperature $T$ is linked to the collision energy and  is  independent of the colliding system. At the LHC, the value of $T$ is well consistent with chiral crossover temperature calculated in LQCD.  The fireball volume parameter   at mid-rapidity  scales linearly with $dN_{ch}/d\eta$,  whereas the canonical  volume parameter,  where strangeness is exactly conserved,  is  larger than the fireball volume  at mid-rapidity. Thus, exact conservation of strangeness is to be imposed in the full phase-space rather than in the experimental acceptance at mid-rapidity. The inclusion  of interactions via phase-shift data improves essentially the fits of the model to proton and hyperon yields   data. 

\section{Acknowledgements}
P.M.L and K.R acknowledge the support by the Polish National Science Center (NCN) under Opus grant no.2018/31/B/ST2/01663.  K.R. also acknowledges partial support of the Polish Ministry of Science and Higher Education. N.S. acknowledges the support of SERB Ramanujan Fellowship (D.O. No. SB/S2/RJN-084/2015) of the Department of Science and Technology, Government of India.

% Our results show some interesting  new features:
%\begin{itemize}
%\item The inclusion  of interactions via phase shift data improves the fits of the HRG model.
%\item The strangeness increase can  be described  by imposing exact strangeness conservation for low multiplicities, the use of two
%volumes improves the fits especially for small charged multiplicities.
%\end{itemize}
%
%In a future publication we intend to address the question of baryon conservation and its consequences.
%%%%%%%%%%%%%%%%%%%%%%%%%%%%%%%%%%%%%%%%%%%%%%%%%%%%%%%%%%%%%%%%%%%%%%%%%%%%

%\bibliographystyle{plainnat}
%\bibliographystyle{plain}
%\bibliographystyle{unsrt}
%\bibliographystyle{apsrev}
%\bibliography{cprs}
%\end{document}

\end{document}